\documentclass[
    ,final            
  ]
  {aipproc}

\layoutstyle{6x9}

\begin{document}

\newcommand{\gein}{$G_{EI}$ }
\newcommand{\gen}{$G_E$ }

\title{Percolation Approach to Study Connectivity in Living Neural Networks}



\author{Jordi Soriano}{
  address={Department of Physics of Complex Systems, Weizmann Institute of Science. Rehovot 76100, Israel}
}
\author{Ilan Breskin}{
  address={Department of Physics of Complex Systems, Weizmann Institute of Science. Rehovot 76100, Israel}
}
\author{Elisha Moses}{
  address={Department of Physics of Complex Systems, Weizmann Institute of Science. Rehovot 76100, Israel}
}
\author{Tsvi Tlusty}{
  address={Department of Physics of Complex Systems, Weizmann Institute of Science. Rehovot 76100, Israel}
}

\begin{abstract}
We study neural connectivity in cultures of rat hippocampal neurons. We measure the
neurons' response to an electric stimulation for gradual lower connectivity, and
characterize the size of the giant cluster in the network. The connectivity undergoes a
percolation transition described by the critical exponent $\beta \simeq 0.65$. We use a
theoretic approach based on bond--percolation on a graph to describe the process of
disintegration of the network and extract its statistical properties. Together with
numerical simulations we show that the connectivity in the neural culture is local,
characterized by a gaussian degree distribution and not a power law one.
\end{abstract}

\maketitle


\section{Introduction}

Neurons in living networks form a highly rich web of connections in which activity flows
between neurons through synapses. The most fascinating living neural network is the human
brain, but its complex architecture, functionality, and computation capability is still
far from being fully understood. More impressive is that the 100 billion neurons are not
randomly connected, but rather form elaborate circuits with specific tasks. Connectivity
thus appears as the fundamental feature to understand the potential of a living neural
network.


Unravelling the detailed connectivity diagram of a living neural network is a painstaking
process. For a brain, a small section of it, or even for a small neural culture, with
$\sim 10^5$ neurons and $\sim 10^7$ connections in just 1 mm$^2$, this task is, at
present, unfeasible. In the brain, substantial progress has been attained in the
description of the connectivity in the mammalian cortex
\cite{Mountcastle-1997,Binzegger-2004}, or the analysis of brain functional networks
\cite{Sporns-2004,Eguiluz-2005}. However, only in the small invertebrate {\it C. elegans}
\cite{White-1986} it has been possible to map out, in a Herculean project, the
connectivity of its 302 neurons. It is not surprising then that other approaches,
different than the pure physiological ones, are being introduced to extract information
about the connectivity of neural networks or, at least, some relevant statistical
properties.

Biological neural networks have caught the attention of Physicists and Mathematicians
following the ``burst'' of interest that complex networks and random graphs have
experienced in the last decade \cite{Newman-2003,Newman-2006}. Graph theory has permitted
to reduce the complexity of a rich variety of natural and artificial networks (e.g.
Internet, e--mail, social, collaborations, or genetic networks) in terms of basic
concepts that retain their most important features, such as the presence of a power law
connectivity, clustering, or the small world phenomena. One of these concepts, which is
related with percolation theory \cite{Stauffer-1991,Callaway-2000}, is the
characterization of the giant cluster (or giant component) of the network and how it
disintegrates as links or nodes are removed. A power law connectivity for instance makes
the network robust to random attacks, but vulnerable to directed attacks, since the
removal of just a small number of highly connected nodes destroys the giant component
\cite{Callaway-2000}. The problem of resilience is of great interest for biological
neural networks, and makes the study of neural connectivity of enormous importance.

Next, we will see how concepts of graph and percolation theory can be used to extract
statistical information about the connectivity in living neural networks. We will
describe our experimental results on connectivity in neural cultures and their analysis
in terms of bond--percolation on a graph \cite{Breskin-2006}. Together with numerical
simulations of the model we show that the connectivity in neural cultures is
characterized by a Gaussian distribution (and not a power law one), and with the presence
of some locality and clustering.

\begin{figure}
\includegraphics[height=.14\textheight]{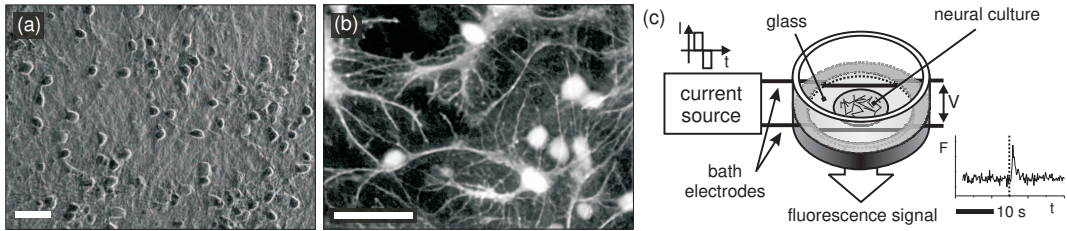}
\caption{(a) Phase contrast image of a small region of a neural culture. Spherical
objects are neurons. (b) Fluorescence image. Bright spots are cell bodies. The scale bar
is 50 $\mu$m in both images. (c) Sketch of the experimental setup. The $F(t)$ plot shows
an example of the fluoresce signal of a spiking neuron as a function of time. The
vertical dashed line indicates the excitation time.}\label{Fig:setup}
\end{figure}

\section{Experimental setup and procedure}

Experiments (see Ref.\ \cite{Breskin-2006} for details) were performed on primary
cultures of rat hippocampal neurons, that are plated on glass coverslips (Fig.\
\ref{Fig:setup}a). Neurons develop dendrites and axons shortly after plating, creating a
dense web of connections in a few days (Fig.\ \ref{Fig:setup}b). Cultures were used
14--20 days after plating, when the network is fully developed and its activity is
governed by the balance between excitatory and inhibitory neurons. About $20\%$ of the
neurons are known to be inhibitory \cite{Marom-2002}.

Neurons were electrically stimulated through bath electrodes (Fig.\ \ref{Fig:setup}c),
and the corresponding voltage drop $V$ measured with an oscilloscope. Neuronal activity
was monitored using fluorescence calcium imaging, and data processed to record the
fluorescence intensity $F$ as a function of time. Neural spiking activity is detected as
a sharp increase of the fluorescence intensity.

The connectivity of the network was gradually weakened by blocking the AMPA glutamate
receptors of excitatory neurons with the receptor antagonist CNQX. We studied the role of
inhibition by either leaving active or blocking the GABA receptors with the corresponding
antagonist bicuculine. For simplicity, we label the network containing both excitatory
and inhibitory neurons by $G_{EI}$, and the network with excitatory neurons only by
$G_E$. The response of the network for a given CNQX concentration was quantified as the
fraction of neurons $\Phi$ that fired in response to the electric stimulation at voltage
$V$. Response curves $\Phi(V)$ were obtained by increasing the stimulation voltage from
$2$ to $6$ V in steps of $0.1 - 0.5$ V. At the end of the experiments, the culture was
washed of CNQX to verify that the initial network connectivity was recovered.

\section{Model}

\begin{figure}
\includegraphics[height=.3\textheight]{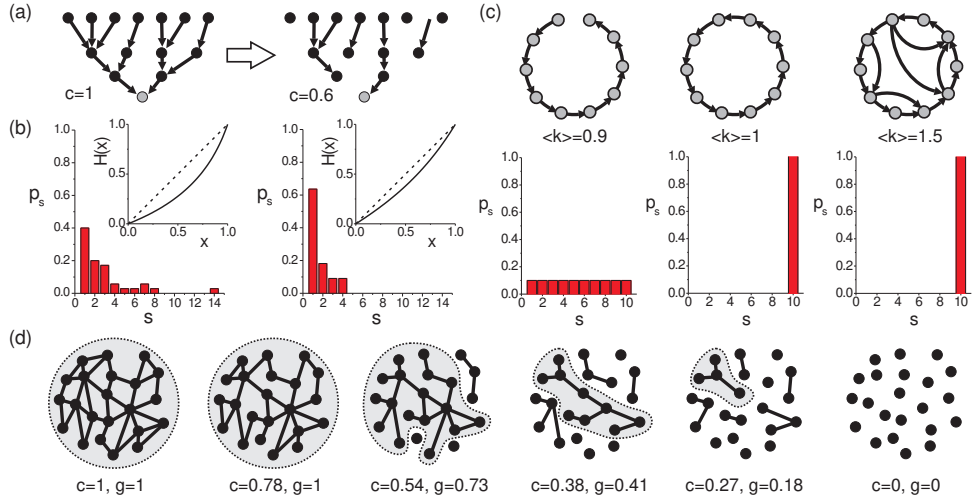}
\caption{(a) Percolation model. The neuron represented in grey fires either in response
to an external excitation or if any of its input neurons fire. At the highest
connectivity, this neuron has input--clusters $s-1=0$ (self--excitation), $7$ (left
branch), $6$ (right branch), and $13$ (both branches). At lower connectivity, its
input--clusters are reduced to sizes $0$ and $3$. (b) Corresponding $p_s(s)$
distributions, obtained by counting all input--clusters for all neurons. Insets: $H(x)$
functions for the $p_s(s)$ distributions (solid lines), compared with independent
neurons, $H(x)=x$ (dashed lines). (c) Example of the sensitivity of $p_s(s)$ to loops.
Left: neurons forming a chain--like connectivity give a $p_s(s)$ distributed uniformly.
Center: closing the loop by adding just one link collapses $p_s(s)$ to a single peak.
Right: additional links increase the average connectivity $\langle k \rangle$, but do not
modify $p_s(s)$. (d) Concept of giant component. The grey areas outline the size of the
giant component $g$ (biggest cluster) for gradually smaller connectivity
$c$.}\label{Fig:model}
\end{figure}

We consider a simplified model of the network in terms of bond--percolation on a graph.
The neural network is represented by the directed graph $G$. Our main simplifying
assumption is the following: \ A neuron has a probability $f=f(V)$ to fire as a direct
response to the electric excitation, and it always fires if any one of its input neurons
fire (Fig.\ \ref{Fig:model}a). This approach ignores the fact that more than one input is
needed to excite a neuron, and that connections are gradually weakened rather than
abruptly removed. However, the aim of the model is to provide the simplest scenario to
understand the experimental observations, and not the actual, highly complex behavior of
neural cultures. $f$ is the natural unit in which to measure the response of the network,
and by a change of variable the measured response curves $\Phi(V)$ can be expressed as
$\Phi(f)$.

The fraction of neurons in the network that fire for a given value of $f$ defines the
firing probability $\Phi (f)$. $\Phi (f)$ increases with the connectivity of $G$, because
any neuron along a directed path of inputs may fire and excite all the neurons downstream
(Fig.\ \ref{Fig:model}a). All the upstream neurons that can thus excite a certain neuron
define its input--cluster or excitation--basin. It is therefore convenient to express the
firing probability as the sum over the probabilities $p_{s}$ of a neuron to have an
input--cluster of size $s-1$ (Figs.\ \ref{Fig:model}b--c),
\begin{eqnarray}
\Phi (f) &\nonumber=&f+(1-f)P\left( \mbox{any input neuron fires}\right)\\
&\label{Eq:Phi}=&f+(1-f)\sum_{s=1}^{\infty }p_{s}\left( 1-\left( 1-f\right) ^{s-1}\right)
=1-\sum_{s=1}^{\infty }p_{s}\left( 1-f\right) ^{s},
\end{eqnarray}
where we used the probability conservation $\sum\nolimits_{s}p_{s}=1$. $\Phi (f)$
increases monotonically with $f$ and ranges between $\Phi (0)=0$ and $\Phi (1)=1$. The
deviation of  $\Phi (f)$ from linearity manifests the connectivity of the network (for
disconnected neurons $\Phi (f)=f$). Eq.\ \eqref{Eq:Phi} indicates that the observed
firing probability $\Phi (f)$ is actually one minus the generating function $H(x)$ (or
the $z$--transform) of the cluster--size probability $p_{s}$ \cite{Shante-1971},
\begin{equation}\label{Eq:Hx}
H(x)=\sum_{s=1}^{\infty }p_{s}x^{s}=1-\Phi (f),
\end{equation}
where $x=1-f$. One can extract from $H(x)$ the input--cluster size probabilities $p_{s}$,
formally by the inverse $z$--transform, or more practically, in the experiments, by
fitting $H(x)$ to a polynomial in $x$.

Once a giant component emerges (Fig.\ \ref{Fig:model}d) the observed firing pattern is
significantly altered. In an infinite network, the giant component always fires no matter
what the firing probability $f$ is. This is because even a very small $f$ is sufficient
to excite one of the infinitely many neurons that belong to the giant component. We
account for this effect by splitting the neuron population into a fraction $g$ that
belongs to the giant component and always fires and the remaining fraction $1-g$ that
belongs to finite clusters. This modifies the summation on cluster sizes into
\begin{equation}
\Phi (f) =g+(1-g)\left[f+(1-f)P\left( \mbox{any inp. neu. fires}\right)\right]
=1-(1-g)\sum_{s=1}^{\infty }p_{s}\left( 1-f\right) ^{s}.
 \end{equation}%
As expected, at the limit of almost no self--excitation $f\rightarrow 0$ only the giant
component fires, $\Phi (0)=g$, and $\Phi (f)$ monotonically increases to $\Phi (1)=1$.
With a giant component present the relation between $H(x)$ and the firing probability
changes, obtaining
\begin{equation}\label{Eq:Hx-giant} H(x)=\sum_{s=1}^{\infty }p_{s}x^{s}=\frac{1-\Phi (f)}{1-g}.
\end{equation}
The size of the giant component decreases with the connectivity. At a critical
connectivity $c_0$ the giant component disintegrates and its size is comparable to the
average cluster size in the neural network. This behavior corresponds to a percolation
transition, separating a system of small, fragmented clusters to one with a fast growing
giant cluster that comprises most of the network.


\section{Experimental results}

\begin{figure}
\includegraphics[height=.37\textheight]{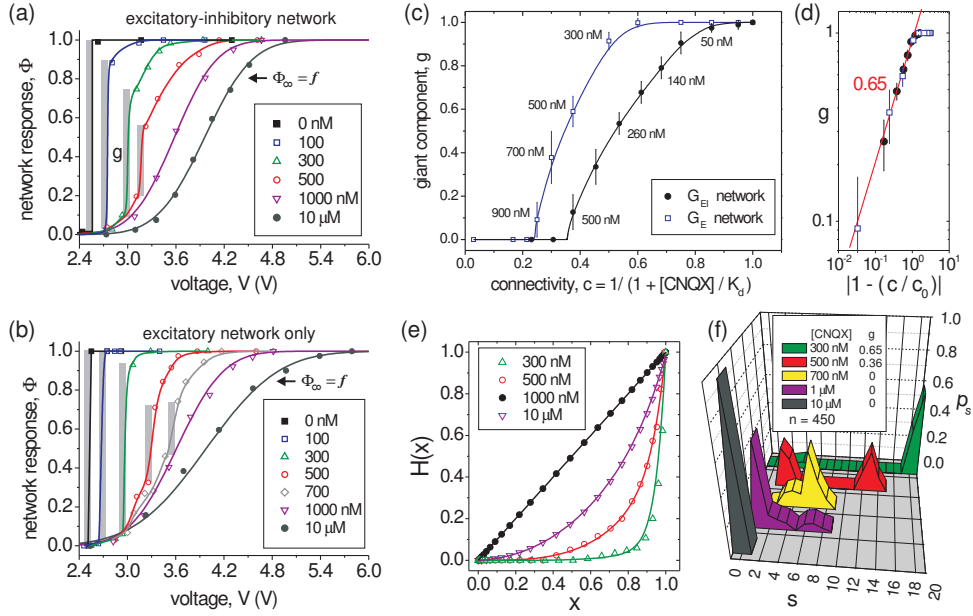}
\caption{(a) and (b) Examples of response curves $\Phi(V)$ for \gein (top) and \gen
(bottom) networks at different concentrations of CNQX. Grey bars indicate the size of the
giant component. Lines are a guide to the eye except for $1$ $\mu$M and $10$ $\mu$M that
are fits to error functions. (c) Size of the giant component as a function of the
connectivity $c$ for \gein networks (circles) and \gen networks (squares). Lines are a
guide to the eye. Some CNQX concentrations are indicated for clarity. (d) Log--log plot
of the power law fits $g \sim |1-c/c_o|^{\beta}$, with $c_o = 0.36 \pm 0.02$, $\beta =
0.66 \pm 0.05$ for \gein, and $c_o = 0.24 \pm 0.02$, $\beta = 0.63 \pm 0.05$ for \gen.
(e) $H(x)$ functions for the response curves shown in (a) and for [CNQX] $> 100$ nM.
Lines are polynomial fits up to order $20$. (f) Corresponding cluster size distribution
$p_s(s)$.}\label{Fig:exp-results}
\end{figure}

Examples of the response curves $\Phi(V)$ for \gein and \gen networks are shown in Figs.\
\ref{Fig:exp-results}a and \ref{Fig:exp-results}b. At one extreme, with [CNQX] $=0$ the
network is fully connected. All neurons form a single cluster that comprises the entire
network. A few neurons with low firing threshold suffice to activate the entire culture,
leading to a very sharp response curve. At the other extreme, with high concentrations of
CNQX ($\simeq$ 10 $\mu$M) the network is completely disconnected, and the response curve
is given by the individual neurons' response. $\Phi(V)$ for individual neurons (denoted
as $\Phi_{\infty}(V)$) is well described by an error function $\Phi(V)=0.5+0.5\;
\mathrm{erf}\left(\frac{V-V_0}{\sqrt 2 \,\sigma_0}\right)$. This indicates that the
firing threshold of a neuron in the network follows a gaussian distribution with mean
$V_0$ and width $2\sigma_0$.

Intermediate CNQX concentrations induce partial blocking of the synapses. Some neurons
break off into separated clusters, while a giant cluster still contains most of the
remaining neurons. The response curves are then characterized by a big jump that
corresponds to the biggest cluster ({\it giant component}), and two tails that correspond
to smaller clusters of neurons with low or high firing threshold. Beyond a critical
concentration (around $500$ nM for \gein networks and $700$ nM for \gen networks) a giant
component cannot be identified and the whole response curve is then also well described
by an error function.

The biggest cluster in the network characterizes the giant component $g$. For each
response curve, $g$ is measured as the biggest fraction of neurons that fire together in
response to the electric excitation, as shown by the grey bars in Figs.\
\ref{Fig:exp-results}a and \ref{Fig:exp-results}b. The size of the giant component was
studied as a function of the connectivity probability (or synaptic strength) between two
neurons \cite{Breskin-2006}, given by $c= 1/(1+[\mathrm{CNQX}]/K_d)$, with $K_d = 300$
nM, and takes values between $0$ (full blocking) and $1$ (full connectivity).

The breakdown of the network for both \gein and \gen networks is shown in Fig.\
\ref{Fig:exp-results}c. The giant component for \gein networks breaks down at much lower
CNQX concentrations compared with \gen networks, indicating that the effect of inhibition
on the network is to effectively reduce the number of inputs that a neuron receives on
average. The behavior of the giant component indicates that the neural network undergoes
a percolation transition, described by the power law $g \sim |1 - c/c_o|^{\beta}$. Power
law fits for \gein and \gen networks give the same $\beta \simeq 0.65$ within the
experimental error (Fig.\ \ref{Fig:exp-results}d), indicating that $\beta$ is an
intrinsic property of the network.

Finally, we have studied the size distribution $p_s(s)$ for clusters that do not belong
to the giant component. $p_s(s)$ has been obtained by constructing the experimental
function $H(x)$ and after fitting a polynomial $\sum\nolimits_{s}p_{s}x^s$. Since $f
\equiv \Phi_{\infty}(V)$ is the response curve for individual neurons (Figs.\
\ref{Fig:exp-results}a and \ref{Fig:exp-results}b) and $x=1-f$, the function $H(x)$ for
each response curve is obtained by plotting $1-\Phi(V)$ as a function of
$1-\Phi_{\infty}(V)$. For curves with a giant component present, its contribution is
eliminated and the resulting curve normalized by the factor $1-g$. Fig.\
\ref{Fig:exp-results}e shows the $H(x)$ functions for the response curves of Fig.\
\ref{Fig:exp-results}a. The corresponding $p_s (s)$ distribution, obtained from fits up
to order $20$, is shown in Fig.\ \ref{Fig:exp-results}f. Overall, the clusters start out
relatively big to rapidly become smaller for gradually higher concentrations of CNQX.
$p_s(s)$ is characterized by isolated peaks, indicating that loops and strong locality
may be present in the neural culture. An example that illustrates the strong effect of
loops on $p_s (s)$ is shown in Fig.\ \ref{Fig:model}c. Since $p_s (s)$ is obtained by
fitting polynomials on $H(x)$, the accuracy in the description of $p_s (s)$ is limited by
the resolution of $H(x)$ which, in turn, is limited by the experimental resolution in
$\Phi(V)$. In addition, since $p_s(s)$ is a probability distribution, the fit is carried
out with two constraints, reducing the freedom of fitting: the $p_s$ coefficients have to
be positive and their sum has to be one. Hence, the $p_s (s)$ distribution presented in
Fig.\ \ref{Fig:exp-results}f shows the correct behavior, but not the precise details of
the distribution of input--clusters.

\section{Numerical simulations}

The model has been derived from classic bond percolation theory and has an analytic
solution that yields precise results. However, the model contains a series of simplifying
assumptions that may have an effect on the results. The numerical simulations that we
present next are oriented to investigate the effect of removing or relaxing these
assumptions, and to provide a physical picture for the connectivity in the network.

Three assumptions of the model are unrealistic. First, it assumes that one input suffices
to activate a neuron, while in reality a number of input neurons must spike for the
target neuron to fire. Second, the effect of CNQX is to bind and block AMPA glutamate
receptor molecules, and consequently to continuously reduce the synaptic strength, so
that bonds are in reality gradually weakened rather than abruptly removed. Third, the
model assumes a tree-like connectivity, while in the living culture loops and clusters
may exits. The numerical simulations have been applied to test that none of these
assumptions change the main results of the model, i.e. that the giant component undergoes
a percolation transition at a critical connectivity $c_0$, and that the analysis of
$H(x)$ provides the distribution of input--clusters in the network.

The numerical simulations also provide the framework to study different degree
distributions and their effect in the critical exponent $\beta$. A Gaussian distribution
gives $\beta \simeq 0.66$, as in the experiments, while a power law distribution, $p_k(k)
\sim k^{-\lambda}$, gives $\beta$ equal to or larger than one, where its exact value
depends on the exponent $\lambda$ \cite{Schwartz-2002}.

\subsection{Numerical method}

The neural network was simulated as a directed random graph $G(N, k_{I/O})$ in which each
vertex is a neuron and each edge is a synaptic connection between two neurons
\cite{Newman-2001}. The graph was generated by assigning to each edge an input/output
connectivity $k_{I/O}$ according to a predetermined degree distribution. Next, a
connectivity matrix $C_{ij}$ was generated by randomly connecting pairs of neurons with a
link of initial weight 1 until each vertex was connected to $k_{I/O}$ links. The process
of gradual weakening of the network was simulated in one case by removing edges, and in
the second case by gradually reducing the bond strength from $1$ to $0$. The connectivity
$c$ is defined for the case of removing bonds as the fraction of remaining edges, and for
the case of weakening bonds as the bond strength.

Each neuron has a threshold $v_i$ to fire in response to the external voltage, and all
neurons have a threshold $T$ to fire in response to the integrated input from their
neighbors. Since the experiments show that the probability distribution for independent
neurons to fire in response to an external voltage is Gaussian, the $v_i$'s are
distributed accordingly. For the simple case of removing links, the global threshold $T$
differentiates networks where a single input suffices to excite a target neuron from
those where multiple inputs are necessary. When links are weakened $T$ plays a more
subtle role, and determines the variable number of input neurons that are necessary to
make a target neuron spike.

The state of each neuron, inactive ($0$) or active ($1$) was kept in a state vector $S$.
In the first simulation step, a neuron fires in response to the external voltage if the
``excitation voltage'' $V$ is greater than its individual threshold $v_i$, i.e. $V\geq
v_i \rightarrow S_i=1$.

In the subsequent simulation steps, a neuron fires due to the internal voltage if the
integration over all its inputs at a given iteration is larger than $T$: $\sum C_{ij}S_j
\geq T \rightarrow S_i=1$. The simulation iterates until no new neurons fire. The network
response $\Phi(V)$ is then measured as the fraction of neurons that fired during the
stimulation. The process is repeated for increasing values of $V$, until the entire
network gets activated, $\Phi(V)=1$. Then, the network is weakened and the exploration in
voltages started again.

\begin{figure}
\includegraphics[height=.365\textheight]{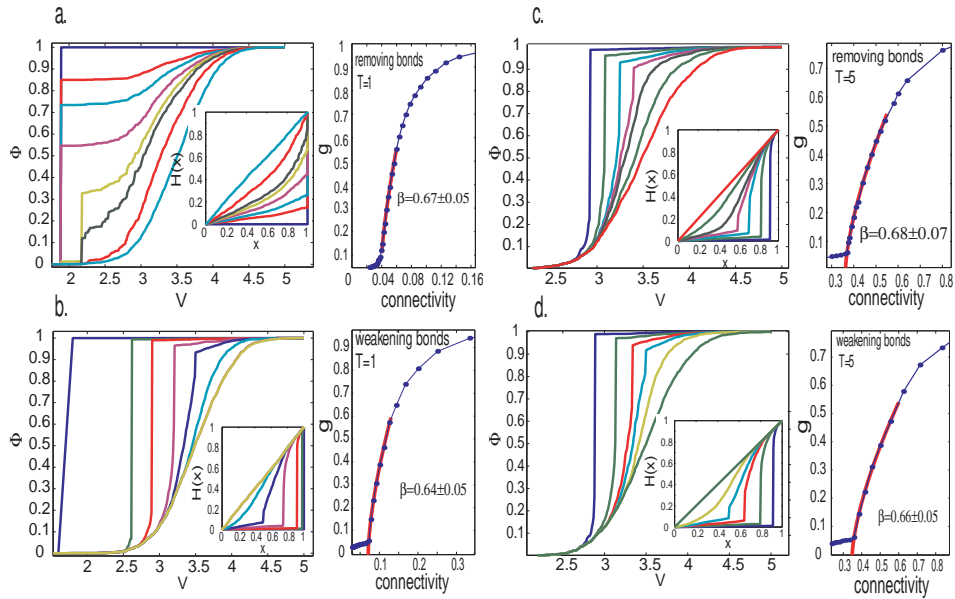}
\caption{Numerical simulations for 4 different cases. Shown are the response curves
$\Phi(V)$, the corresponding $H(x)$ functions (inset), and the characterization of the
percolation transition for (a) removing edges, T=1; (b) weakening edges, T=1; (c)
removing edges, T=5; and (d) weakening edges, T=5.}\label{Fig:simulations}
\end{figure}

\subsection{Simulations results}

\subsubsection{Analysis of the model}

To study the validity of the model, we have first considered 4 different situations:
removing or weakening edges, and for $T=1$ or $T=5$. In all cases the connectivity is set
to be Gaussian for both input and output degree distributions. The results of the
simulation are presented in Fig.\ \ref{Fig:simulations}. All 4 studied cases give
qualitatively similar results, with response curves $\Phi(V)$ that are comparable to the
ones observed experimentally, and with a giant component clearly identifiable. The
analysis of the percolation transition gives $\beta \simeq 0.66$ in all 4 cases, in
agreement with the value measured experimentally. As expected, the simulations with
weakening bonds and for $T=5$ (five spiking neurons required to excite the target neuron)
provide the response curves that are more similar to the ones observed experimentally.
However, it is remarkable that the simplest case of the model (breaking bonds with $T=1$)
already gives valid results. This indicates that, with the limitations of the model, the
percolation approach proves to be remarkably powerful in describing the behavior observed
experimentally.

The other important assumption of the model is the effect of the presence of loops in the
network. Although loops are very rare in a random (Erd\"{o}s-R\'{e}nyi) graph, the
connectivity in neural cultures is not random, and locality and neighboring probably may
play an important role. However, graph theory tells us that most loops will be found in
the giant component, where all neurons anyway light up and their effect is therefore
irrelevant to our analysis. Clusters outside the giant component are in general
tree--like, and thus the important analysis to be considered is what happens when finite
clusters do have loops. The simulations show that the response curves and the percolation
transition are not significantly altered if loops are allowed, providing similar results
to the ones shown in Fig.\ \ref{Fig:simulations}.

Loops, however, do affect the clusters size distribution $p_s(s)$, which is then
characterized by the presence of isolated peaks. To explore to which extent $p_s(s)$ was
sensitive to loops, we performed simulations considering different levels of clustering.
The first graph that we analyzed was one with an artificially induced highly clustered
connectivity. We generated a network where most of the links are located in highly
connected clusters, with only weak connections between clusters. The $p_s(s)$
distribution obtained from the breakdown of the connectivity in such a clustered network
showed that the position of the dominant peaks corresponded to the size of the highly
connected clusters. Next, having demonstrated the importance of highly connected
clusters, we went on to consider realizations of the graph that would be more similar to
the experimental network. To do that, we introduced the notion of geometry and of
distance, placing all vertices on a spatial grid. Three different configurations were
used: (i) a Gaussian connectivity with no locality, (ii) a Gaussian connectivity with
local connections and (iii) a Gaussian connectivity with distance dependent link
strength. For the first case no dominant peaks where identifiable. For the second one the
existence of isolated peaks is more apparent. But for the third case the reinforced
connectivity significantly increases the probability to have isolated input--clusters,
similar to what we observe experimentally.

\subsubsection{Role of inhibition and analysis of H(x)}

We have studied the role of inhibition in the network by randomly selecting a subgroup of
nodes and assigning them negative weights to simulate inhibitory neurons. Then,
simulations with the same conditions described above were repeated and different
excitation/inhibition ratios explored. The results indicate that the critical exponent
$\beta$ is independent of the balance between excitation and inhibition, in agreement
with the experimental observations. The results also show that the critical connectivity
$c_0$ at which the giant component disintegrates does depend on the number of inhibitory
neurons, and that this is a linear dependence.

Finally, we have verified with the simulations that the cluster distribution $p_s(s)$
obtained from the polynomial fit of $H(x)$ does not differ significantly from the
$p_s(s)$ distribution directly extracted from the connectivity matrix $C_{ij}$. Small
deviations are a consequence of the constraints $\sum p_s=1$ and $0 \leq p_s \leq 1$ in
the polynomial fits, and in the uncertainty in removing the contribution of the giant
component in the $H(x)$ functions. This analysis gives validity to the $p_s(s)$
distribution measured experimentally.

\section{Discussion and conclusions}

By comparing the exponent $\beta$ measured experimentally with the one obtained from the
simulations we conclude that the connectivity in the neural culture is Gaussian.
Simulations, however, are based on a random graph, while the real neural network is not,
and one may think that the neural culture is actually better described by a
two--dimensional, lattice--like network. Percolation on two--dimensional lattices gives a
critical exponent $\beta \simeq 0.14$, independent on the lattice structure. The value of
the exponent increases rapidly with the dimensionality of the lattice, with $\beta \simeq
0.41$ and $0.64$ for three and four dimensions, respectively. In a system described by a
2--D structure, additional dimensions can be viewed as a gradual increase of long--range
correlations.

The physical picture that we think may exist in the neural culture is that neurons are
essentially connected to their neighbors, but with some long--range correlations. Axons
can easily extend $300$ $\mu$m in a neural culture, connecting neurons as far as $30$
cell bodies. The concept that locality is important is in fact quite natural when one
thinks of the nature of the culture. Neurons are distributed homogeneously over the
glass, and most likely all neurons start to form connections at the same time and at the
same rate. This hints at a structure where neurons are highly connected with their
neighbors. This is also suggested by the distribution of input--clusters $p_s(s)$, which
shows that neurons are highly connected between them even after the giant component has
begun disintegrating, forming local clusters with a significant presence of loops. We
have also seen that neurons surrounded by many others tend to fire first in response to
the external excitation, and that aggregates of neurons tend to fire together, with their
collective response maintained even when the connectivity is reduced.

In summary, we have presented experimental results on the connectivity in neural
cultures, and showed that connectivity undergoes a percolation transition characterized
by a critical exponent $\beta \simeq 0.65$. The experimental results were studied in the
framework of percolation on a graph, and extracted the distribution of connected
components in the network. Numerical simulations of the model were used to construct a
physical picture of the connectivity in the neural network, and showed that the
connectivity is characterized by a Gaussian degree distribution, with strong locality and
clusterization.


\begin{theacknowledgments}
We thank L. Gruendlinger, M. Segal, J.-P. Eckmann, and O. Feinerman for their insight. J.
S. acknowledges the financial support European Training Network PHYNECS, project No.
HPRN-CT-2002-00312. Work supported by the Israel Science Foundation, grant 993/05, and
the Minerva Foundation, Munich, Germany.
\end{theacknowledgments}

\bibliographystyle{aipproc}   


\begin{thebibliography}{9}

\bibitem{Mountcastle-1997}
V.~B. Mountcastle, \emph{Brain} \textbf{120}, 701 (1997).

\bibitem{Binzegger-2004}
T. Binzegger, R.~J. Douglas, and K.~A.~C. Martin, \emph{J. Neurosci.} \textbf{24}, 8441
(2004).

\bibitem{Sporns-2004} O. Sporns, D.~R. Chialvo, M. Kaiser, and C.~C. Hilgetag, \emph{Trends Cogn. Sci.} \textbf{8}, 418 (2004).

\bibitem{Eguiluz-2005}
V.~M. Egu{\'i}luz, D.~R. Chialvo, G.~A. Cecchi, M.~Baliki, and A.~V. Apkarian,
\emph{Phys. Rev. Lett.} \textbf{94}, 018102 (2005).

\bibitem{White-1986} J.~G. White, E. Southgate, J.~N. Thomson, and S. Brenner, Phil. Trans. R. Soc. Lond. B \textbf{314}, 340 (1986).

\bibitem{Newman-2003} M.~E.~J. Newman, \emph{The structure and function of complex networks}, SIAM Review
\textbf{45}, 167 (2003).

\bibitem{Newman-2006} M.~E.~J. Newman, A.~L. Barab\'asi, D.~J. Watts, \emph{The Structure and Dynamics of
Networks}. (Princeton University Press, 2006).

\bibitem{Stauffer-1991}
D. Stauffer and A. Aharony, \emph{Introduction to Percolation Theory}, 2nd Ed. (Taylor \&
Francis, London, 1991).


\bibitem{Callaway-2000}
D.~S. Callaway, M.~E.~J. Newman, S.~H. Strogatz, D.~J. Watts, \emph{Phys. Rev. Lett.}
\textbf{85}, 5468 (2000).


\bibitem{Breskin-2006}
I. Breskin, J. Soriano, E. Moses, T. Tlusty, \emph{Phys. Rev. Lett.}, in press (October
2006).

\bibitem{Marom-2002} S. Marom and G. Shahaf. Q. Rev. Biophys. {\bf 35},63 (2002).

\bibitem{Shante-1971} F. Harary and G.E. Uhlenbeck. Proc. Nat.
Acad. Sci. {\bf 39} 315 (1952);
 V.K.S. Shante and S. Kirkpatrick. Adv. Phys.
{\bf 20}, 325 (1971).

\bibitem{Schwartz-2002}
N. Schwartz, R. Cohen, D. ben-Avraham, A.-L. Barab\'asi, S. Havlin, \emph{Phys. Rev. E}
\textbf{66}, 015104(R) (2002).

\bibitem{Newman-2001}
M.~E.~J. Newman, S.~H. Strogatz, D.~J. Watts, \emph{Phys. Rev. E} \textbf{64}, 026118
(2001).


%
%

\end{thebibliography}


\end{document}